\documentclass{article}

\usepackage[english]{babel}

\usepackage[letterpaper,top=2cm,bottom=2cm,left=3cm,right=3cm,marginparwidth=1.75cm]{geometry}

\usepackage{amssymb}
\usepackage{amsmath}
\usepackage{graphicx}
\usepackage[colorlinks=true, allcolors=blue]{hyperref}
\usepackage{cleveref}

\title{\textbf{Rock Guitar Tablature Generation via \\Natural Language Processing}}
\author{\textbf{Josue Casco-Rodriguez} \\
Rice University\\
Houston, TX, USA \\
\texttt{jc135@rice.edu}
}
\date{}

\begin{document}
\maketitle

\begin{abstract}
Deep learning has recently empowered and democratized generative modeling of images and text \cite{dall-e2, gpt3}, with additional concurrent works exploring the possibility of generating more complex forms of data, such as audio \cite{its_raw, jukebox}. However, the high dimensionality, long-range dependencies, and lack of standardized datasets currently makes generative modeling of audio and music very challenging. We propose to model music as a series of discrete notes upon which we can use autoregressive natural language processing techniques for successful generative modeling. While previous works used similar pipelines on data such as sheet music and MIDI \cite{musenet, symbolic_music_generation}, we aim to extend such approaches to the under-studied medium of guitar tablature. Specifically, we develop the first work to our knowledge that models one specific genre---heavy rock---as guitar tablature. Unlike other works in guitar tablature generation, we have a freely available public demo at \href{https://huggingface.co/spaces/josuelmet/Metal_Music_Interpolator}{https://huggingface.co/spaces/josuelmet/Metal\_Music\_Interpolator}.\footnote{Our source code is used to train final demo model is available at \href{https://github.com/Josuelmet/Metal-Music-Interpolator}{https://github.com/Josuelmet/Metal-Music-Interpolator}.}
\end{abstract}

\section{Introduction}

\begin{figure}
\centering
\includegraphics[width=\textwidth]{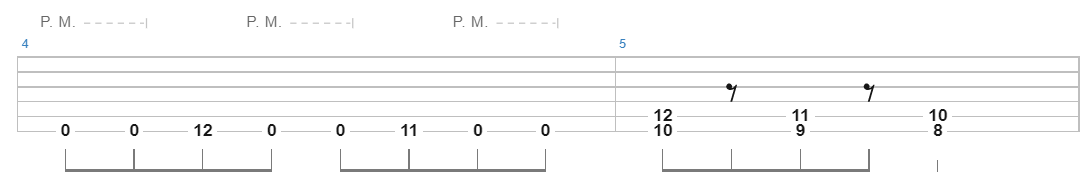}
\caption{\label{fig:tab1} A guitar tablature snippet of two measures written in 4/4 time. Each note is represented not as its pitch, but rather as the specific fret index upon which a player should press upon a certain string so as to produce the note. The fret index of a note on the string it is played on is equivalent to the number of semitones between the note's pitch and the lowest pitch that the string can play. Note that certain notes contain information about their dynamics: for example, the P.M. symbol indicates that certain notes should be played in a semi-muted fashion.}
\end{figure}

\begin{figure}
\centering
\includegraphics[width=0.5\textwidth]{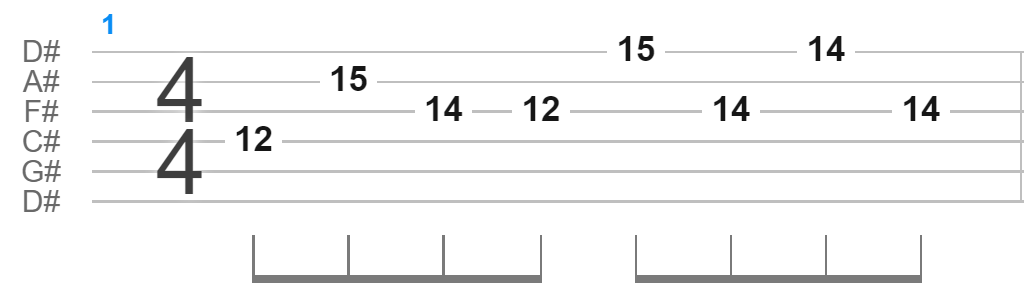}
\caption{\label{fig:tab2} Another guitar tablature snippet in 4/4 time, this time consisting of the iconic first measure of \textit{Sweet Child O' Mine} by Guns N' Roses. The six pitches arranged vertically on the left are the lowest pitches that each of the six guitar strings can play.}
\end{figure}

Music, like images and language, is a fundamental form of art and a quintessential piece of the human experience. Despite the fact that text-to-image models \cite{stable_diffusion, dall-e2, muse, dreambooth, boomerang} have produced explosive breakthroughs in the generation and modeling of visual art, such breakthroughs for music production have not yet been realized; however, recent works, such as OpenAI’s Jukebox \cite{jukebox}, have made progress towards advanced music generation. A large factor in why such breakthroughs have yet to be is that music is challenging to model, requiring sequence-modeling of data mediums that are not as well-understood or intuitive as images or words.
\\
\\
When investigating how a machine can learn to understand or generate music, one can begin by understanding how people learn to work with music. Although music exists as a continuous-time audio signal, people most efficiently understand and analyze music as a pattern of discrete frequencies (for example, the note “A” = 440 Hz) that are played for discretely quantized intervals of time (e.g., quarter-, half-, and whole-notes). As such, sequence-modeling techniques for understanding sequences of discrete data can be leveraged towards music modeling, given a sufficient dataset of music samples represented in discrete forms.
\\
\\
While sheet music and Musical Instrument Digital Interface (MIDI) files are conventional forms of discretely compressed representations of music, one prominent form of music representation that has been studied less is guitar tablature (see \cref{fig:tab1} and \cref{fig:tab2}). While appearing similar to traditional sheet music, guitar tablature differs by representing notes as fret and string indices upon which the instrument-players must place their fingers so as to produce a specific note, since stringed instruments are unique in that most notes that can be played on them have more than one fret and string combination that can produce them.
\\
\\
We develop a new dataset of compressed representations of guitar tablature from one specific genre of music (heavy rock), as well as a neural network architecture that can leverage sequence modeling (such as in long short-term memory networks or natural language processing models) to produce new guitar tablature sequences when conditioned on a brief snippet of an existing guitar tablature. Specifically, the proposed autoregressive model aims to estimate the most likely new tablature token $\mathbf{x}_{N+1}$ when given the previous tokens $\mathbf{x}_1, \mathbf{x}_2, ..., \mathbf{x}_N$ (i.e., estimate the conditional probability $p(\mathbf{x}_{N+1} | \mathbf{x}_1, ..., \mathbf{x}_N)$), thus enabling an iterative procedure through which an $M$-token sequence can be generated from an $N$-token sequence, for $M > N$.

\section{Background}

\subsection{Sequence Models}

\textbf{Recurrent networks.} Sequence modeling is a long-standing problem in machine learning and statistics, with one of its earliest prominent efforts being recurrent neural networks \cite{rnn}. While recurrent neural networks are able to leverage a form of memory to model sequences of theoretically unbounded contexts, in practice they and their recent variants \cite{lstm} struggle to do so, in part due to gradient propagation issues \cite{projunn}. 
\\
\\
\textbf{Transformers.} Meanwhile, transformers \cite{attention_is_all_you_need} circumvent the problems with recurrent neural networks by replacing recurrent operations with one feedforward attention operation that compares every element of a sequence with every other element of the sequence; such an approach could initially seem disadvantaged due to the memoryless nature, inherently finite bounded context \cite{s4}, and $O(N^2)$ runtime of an attention mechanism on a sequence of length $N$ \cite{fmmformer}. However, when combined with additional innovations such as positional embeddings and token dimensionality reduction via vectorization, transformer architectures have yielded enormous advances in sequence and image modeling \cite{gpt3, dall-e2, vit}.
\\
\\
\textbf{Self-attention.} The key behind any transformer architecture is the self-attention mechanism. Given a vector-valued input sequence 
$\mathbf{X} = \left[ \mathbf{x}_1, \mathbf{x}_2, ..., \mathbf{x}_N \right] \in \mathbb{R}^{N \times D_x}$ 
such that each element is $D_x$-dimensional and the transformer feedforward dimension is $D$, a self-attention head transforms $\mathbf{X}$ into an output sequence $\mathbf{\hat{V}}$ through the following: \cite{fmmformer}
\begin{enumerate}
    \item Using the weights $\mathbf{W}_Q, \mathbf{W}_K \in \mathbb{R}^{D \times D_x}$ and $\mathbf{W}_V \in \mathbb{R}^{D_v \times D_x}$, project $\mathbf{X}$ into three distinct matrices---the query, key, and value matrices $\mathbf{Q}, \mathbf{K},$ and $\mathbf{V}$---via these linear transformations:
    \begin{align*}
    \mathbf{Q} &= \mathbf{XW}^T_Q \\
    \mathbf{K} &= \mathbf{XW}^T_K \\
    \mathbf{V} &= \mathbf{XW}^T_V
    \end{align*}
    \item Let us express the query, key, and value matrices as $\mathbf{Q} = \left[ \mathbf{q}_1, ..., \mathbf{q}_N \right]^T, \mathbf{K} = \left[ \mathbf{k}_1, ..., \mathbf{k}_N \right]^T,$ and $\mathbf{V} = \left[ \mathbf{v}_1, ..., \mathbf{v}_N \right]^T$, where the vectors $\mathbf{q}_i, \mathbf{k}_i, \mathbf{v}_i$ for $i \in \{ 1, 2, ..., N \}$ are the query, key, and value vectors, respectively.
    
    Each output sequence vector $\mathbf{\hat{v}}_i$ is calculated by multiplying each value vector $\mathbf{v}_j$ by a score determined as the similarity between the query vector $\mathbf{q}_i$ the key vector $\mathbf{k}_j$ :
    \begin{equation*}
        \mathbf{\hat{v}}_i = \sum_{j=1}^N \text{softmax} \left( \frac{\mathbf{q}_i^T \mathbf{k}_j}{\sqrt{D}} \right) \mathbf{v}_j
    \end{equation*}
    Calculation of $\mathbf{\hat{V}} = \left[ \mathbf{\hat{v}}_1, ..., \mathbf{\hat{v}}_N \right]^T$ can thus be simply expressed as:
    \begin{equation*}
        \mathbf{\hat{V}} = \text{softmax} \left( \frac{\mathbf{QK}^T}{\sqrt{D}} \right) \mathbf{V} = \mathbf{AV},
    \end{equation*}
    where the attention matrix $\mathbf{A}$ is computed by applying the softmax operation to each row of the matrix $\mathbf{QK}^T / \sqrt{D}$ \cite{fmmformer}.
\end{enumerate}

\subsection{Related Works}

\textbf{Music/audio generation.} Our key contribution to musical sequence modeling is publicly available guitar tablature modeling of heavy rock. Various previous and ongoing works have approached music generation both continuous and discrete data modalities. For example, the recent SaShiMi \cite{its_raw} and Jukebox \cite{jukebox} architectures approach audio and music generation in the spaces of continuous waveforms and discrete notes, respectively. The advent of diffusion models \cite{ddpm} has also found influence in a new model combining spectrogram and MIDI music generation \cite{multi-instrument_synthesis}.
\\
\\
\textbf{Guitar tablature literature.} The field of guitar tablature analysis is small but growing, with various works tackling challenges such as graph-based solo analysis \cite{solos_as_networks}, transcription \cite{transcription}, dataset collection, and sequence modeling \cite{animetab, dadagp, groove_modeling}. Of particular importance to our work are AnimeTab \cite{animetab} and DadaGP \cite{dadagp}, since they also opt for a transformer-based approach to statistically generate sequences of guitar tablature. Unlike DadaGP, our model has token representations that are much more simple and easy to understand, is trained on one specific genre, and has a publicly available demo. While our model may share some similarities with AnimeTab, which was published during the development of this work and is supposed to have a demo released soon, our model has a demo already available and is trained on the genre of heavy rock music instead of anime/video game music.

\section{Methods}

\subsection{Data Processing}

\begin{figure}
\centering
\includegraphics[width=0.9\textwidth]{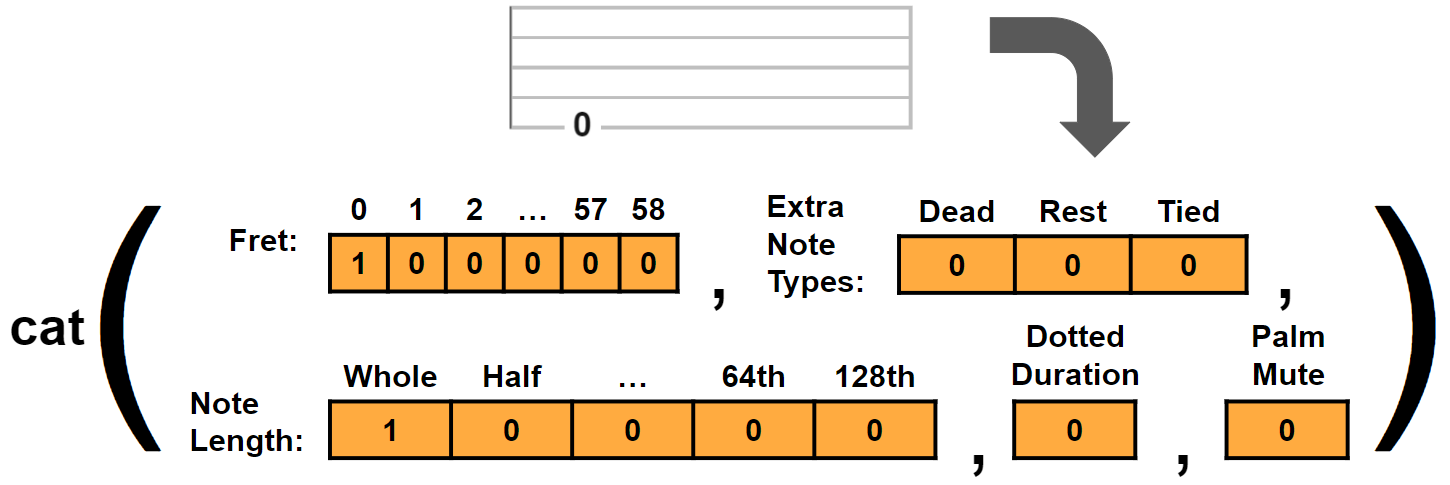}
\caption{\label{fig:embedding} Note embedding scheme illustrated for the example note of a whole note on fret 0 of the lowest string on a guitar/bass. The fret value is one-hot encoded as 0, the note length is one-hot encoded as a whole note, and none of the flags are set to 1 because the note is neither dotted nor palm-muted nor a dead/rest/tied note.}
\end{figure}

\textbf{Initial preprocessing.} The success of statistical inference methods often reflects the quality of data used for training—data preprocessing is just as important to a successful model as the model itself. Our data preprocessing pipeline begins by first collecting a sizeable volume of songs, in guitar tablature format, that accurately represent one subgenre of music\footnote{Complete list of songs: \href{https://github.com/Josuelmet/Metal-Music-Interpolator/blob/main/songs/README.md}{https://github.com/Josuelmet/Metal-Music-Interpolator/blob/main/songs/README.md}}. For each tablature file, every song is first converted into 4/4 time for ease of processing, and is then converted into a Python object via PyGuitarPro\footnote{\href{https://github.com/Perlence/PyGuitarPro}{https://github.com/Perlence/PyGuitarPro}} for ease of querying. Each track of each song (i.e., each instrument or voice of each song, not including drums) is then converted into a one-dimensional list containing each note in the song; each note is represented as a tuple containing the note's pitch (with special designations for tied, dead\footnote{Dead notes are noted that are heavily muted such that they lose a distinct sense of pitch.}, and rest notes) , duration, the chordal nature if applicable (with the represented chords being 4th, diminished 5th, and perfect 5th chords), and two flags indicating whether the note is dotted and whether the note is muted. Note that the pitch of each note is represented not as the musical pitch of each note (e.g., ``A4" or ``C3"), but rather as the fret on the guitar (or bass) upon which a player should place their finger so as to generate the note. Once all songs' notes have been represented as tuples, each tuple is converted to an integer via an invertible dictionary map.
\\
\\
\textbf{Embedding initialization.} After initial pre-processing, each song exists as a set of sequences, where each sequence represents one voice or instrument and contains integers that represent each note. While a naïve sequence model could attempt inference upon these scalar sequences, modern sequence models have found success in instead representing the individual tokens or elements of a sequence as vectors, allowing for more expressive and informative representations token modalities. Unlike previous works \cite{dadagp}, we opt for a simple, but effective, initial token vectorization illustrated in \cref{fig:embedding}. Each initial vectorized token embedding, before training, has 72 dimensions: the first 59 are reserved for one-hot encoding the number of semitones (equivalent to the number of frets on a guitar or bass) between the pitch value and the lowest pitch playable by the given instrument; the next 3 dimensions are flags indicating if the note is a dead, rest, or tied note; the next 8 dimensions one-hot encode the note's duration (e.g., whole-, half-, and quarter-notes); and the last two dimensions are flags indicating if the note's duration is dotted and if the note is played in a muted fashion. While the vectorized token embeddings are trained, and thus iteravely refined, during the training process, we found that our hand-crafted initialization scheme performed better than default random token initialization. As in any other successful transformer architecture, each token also has a positional embedding that accompanies the vectorized token embedding before going into the transformer model.

\subsection{Model Architecture}
In addition to implementing the entire data preprocessing pipeline with the only starting point being PyGuitarPro for querying tablature files as Python objects, we also had to manually implement the transformer architecture, since we opted for a mini-GPT model due to the relatively small dataset at our disposal compared to the number of sequences and number of parameters used to train conventional language models \cite{gpt3}. Since we were not using a pre-written transformer model, we also had to implement a causal masking mechanism to ensure that the transformer cannot use information from any tokens after the token it is trying to predict. After extensive hyperparameter tuning on a 90-10 testing-validation, split, our final hyperparameter values are located in \cref{tab:hyperparameters}. The final mini-GPT architecture consists of an embedding layer (as described earlier), three transformer blocks in sequence, and one final feedforward layer that returns a 1629-dimensional vector representing the conditional probability of the next token's value $p \left( \mathbf{x}_{N+1} | \mathbf{x}_1, ..., \mathbf{x}_N \right)$.

\begin{table}
\centering
\begin{tabular}{l|r}
Hyperparameter & Value \\\hline
N (sequence length) & 100 \\
Output dimensionality (number of unique tokens) & 1629 \\
Initial embedding dimensionality & 72 \\
Number of transformer blocks & 3 \\
Transformer feedforward dimension & 512 \\
Transformer dropout rate & 30\% \\
Attention heads per transformer & 10 \\
Initial learning rate & 0.003 \\
$\beta_1$ (Adam parameter) & 0.96 \\
Batch size & 512 \\
Training epochs (determined by early stopping) & $\sim120$
\end{tabular}
\caption{\label{tab:hyperparameters} Hyperparameters of the final mini-GPT model.}
\end{table}

\section{Results}

\begin{figure}
\centering
\includegraphics[width=\textwidth]{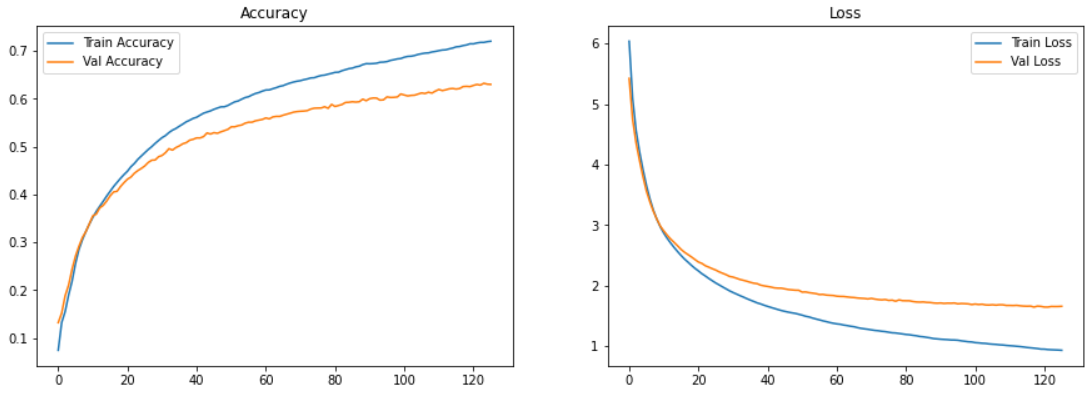}
\caption{\label{fig:loss} Training and validation accuracy and loss of the final mini-GPT model using a 90-10 training-validation split.}
\end{figure}

\textbf{Performance.} Using the hyperparameters specified in \cref{tab:hyperparameters} and early stopping based on the validation loss, our mini-GPT model achieved over 70\% training accuracy and over 60\% validation accuracy, as shown in \cref{fig:loss}. While it was possible to improve the training accuracy by reducing the transformer dropout, we empirically found that doing so worsened validation generalization due to overfitting. Qualititative evaluation of our model is made possible through the interactive demo provided\footnote{\href{https://huggingface.co/spaces/josuelmet/Metal_Music_Interpolator}{https://huggingface.co/spaces/josuelmet/Metal\_Music\_Interpolator}}. Not only is our work the first, to our knowledge, that provides a publicly accessible generation demo, but our model's success also further validates the usage of and need for natural language processing techniques in the realm of audio and music generation.
\\
\\
\textbf{Future works.} Straightforward and interesting extensions of our model include more explicity incorporating rhythmic information \cite{groove_modeling} in the embeddings or as a separate embedding; such an approach would be better able to capture and understand the differences between notes that land on downbeats, upbeats, and backbeats. Another extension would be to model multiple tracks or voices at a time, allowing for modeling of rich chords or drum sequences. For note-based chord or drum modeling, some form of specialized translational invariance could be key to ensuring that the autoregressive model understands that, when multiple notes are placed at the same time, their order in a sequence does not matter. Using a more novel attention mechanism \cite{fmmformer} could further improve performance. Diffusion models could also play a role in generating sequences of guitar tablature, but whether they can outperform autoregressive language models has yet to be shown; recent work has shown that transformers and diffusion models can be combined to produce state-of-the-art results in music generation \cite{multi-instrument_synthesis}.

\bibliographystyle{ieeetr}
\bibliography{main}

\end{document}